\documentclass[aps,prd,secnumarabic,nobibnotes,twocolumn,superscriptaddress]{revtex4-1}
\usepackage{amsfonts}
\usepackage{mathrsfs}
\usepackage{amsmath}
\usepackage{color}
\usepackage{natbib}
\usepackage{graphicx}
\usepackage{bm}
\usepackage{amssymb}
\usepackage{xspace}
\usepackage{epstopdf}
\usepackage{dcolumn}
\usepackage{multirow}
\usepackage[colorlinks=true, letterpaper=true, pdfstartview=FitV, linkcolor=blue, citecolor=blue, urlcolor=blue]{hyperref}
\usepackage{wrapfig}

\makeatletter

\newcommand{\Rmnum}[1]{\expandafter\@slowromancap\romannumeral #1@}
\makeatother

\begin{document}
\title{Nodal loop and nodal surface states in Ti$_3$Al family materials}

\author{Xiaoming Zhang}\email{zhangxiaoming87@hebut.edu.cn}
\address{School of Materials Science and Engineering, Hebei University of Technology, Tianjin 300130, China.}
\address{Research Laboratory for Quantum Materials, Singapore University of Technology and Design, Singapore 487372, Singapore}

\author{Zhi-Ming Yu}
\address{Research Laboratory for Quantum Materials, Singapore University of Technology and Design, Singapore 487372, Singapore}

\author{Ziming Zhu}
\address{Research Laboratory for Quantum Materials, Singapore University of Technology and Design, Singapore 487372, Singapore}

\author{Weikang Wu}
\address{Research Laboratory for Quantum Materials, Singapore University of Technology and Design, Singapore 487372, Singapore}

\author{Shan-Shan Wang}
\address{Research Laboratory for Quantum Materials, Singapore University of Technology and Design, Singapore 487372, Singapore}

\author{Xian-Lei Sheng}
\address{Department of Applied Physics, Key Laboratory of Micro-nano Measurement-Manipulation and Physics (Ministry of Education),
Beihang University, Beijing 100191, China}

\author{Shengyuan A. Yang}\email{shengyuan\_yang@sutd.edu.sg}
\address{Research Laboratory for Quantum Materials, Singapore University of Technology and Design, Singapore 487372, Singapore}

\begin{abstract}
Topological metals and semimetals are new states of matter which attract great interest in current research. Here, based on first-principles calculations and symmetry analysis, we propose that the family of titanium-based compounds Ti$_3X$ ($X=$Al, Ga, Sn, Pb) are unexplored topological semimetals. These materials feature the coexistence of a nodal loop and a nodal surface in their low-energy band structure.
Taking Ti$_3$Al as an example, we show that the material has an almost ideal nodal loop in the sense that the loop is close to the Fermi level and it is nearly flat in energy with energy variation $<0.25$ meV. The loop is protected by either one of the two independent symmetries: the combined spacetime inversion symmetry and the mirror reflection symmetry. The nodal surface at the $k_z=\pi$ plane is guaranteed by the nonsymmorphic screw rotational symmetry and the time reversal symmetry. We discuss the effect of spin-orbit coupling and construct an effective model for describing the nodal loop. Our findings indicate that the Ti$_3$Al family compounds can serve as an excellent material platform for studying new topological phases and particularly the interplay between nodal-loop and nodal-surface fermions.
\end{abstract}
\maketitle

\section{Introduction}
Crystalline solids may have nontrivial topology encoded in their electronic band structure. The exploration of this notion gives rise to the fascinating field of topological materials that attracts tremendous interest from physics, chemistry, and materials science. Initially studied in the context of insulating states~\cite{add1,add2}, now the concept of band topology has been extended to the metal and semimetal states~\cite{add3,add4,add5}. In these topological metals, topology is associated with the nontrivial band crossings at low energy. For example, in a Weyl semimetal~\cite{add6,add7}, the conduction and valence bands cross at isolated nodal points, around which the band dispersion is linear in all directions and the low-energy electrons behave like Weyl fermions in high-energy physics~\cite{add8}. The Weyl nodal point, acting like monopoles in the $k$-space, is topological robust against weak perturbation. With proper crystalline symmetry, superposing two Weyl points together can make a stable Dirac nodal point, leading to a Dirac semimetal state~\cite{add9,add10,add11,add2nd1}.

For a three-dimensional (3D) system, according to the dimensionality of the band crossing, besides the various 0D nodal points, we can in principle also have 1D nodal lines and 2D nodal surfaces. There have been several nodal-line materials proposed so far~\cite{add12,add13,add14,add15,add16,add17,add18,add19,add20,add21,add22,add23,add24,add25}, showing rich properties. Depending on the shape, the lines may take various forms including a single ring, multiple crossing rings~\cite{add12,add26}, loops traversing the Brillouin zone (BZ)~\cite{add15,add27}, extended chains~\cite{add28,add29}, Hopf links~\cite{add30,add31,add32,add33,add34}, nodal boxes~\cite{add35}, and etc. The lines can also be classified based on the type of band dispersion around the crossing, which leads to the type-I, type-II, and hybrid nodal lines~\cite{add27,add36}. Meanwhile, the possibility of nodal-surface semimetals was just noticed recently, initially by Zhong \emph{et al.}~\cite{add37} in a family of carbon network materials and by Liang \emph{et al.}~\cite{add38} in the BaVS$_3$ family materials. Wu \emph{et al.}~\cite{add39} have proposed two classes of the nodal surfaces, and further provided sufficient conditions and examples for robust nodal surfaces in the presence of spin-orbit coupling (SOC) and in magnetic materials. Compared with nodal-point and nodal-line materials, the proposed nodal-surface materials are much less, and there is no experimental verification of a nodal-surface semimetal yet.

These topological states are expected to exhibit interesting features in their electronic, transport, optic, and magnetic properties. To study these properties, from materials point of view, an important condition is that the nontrivial band crossing must be close to the Fermi level because most electronic properties are determined by the low-energy electrons around the Fermi level. In addition, for experimental studies as well as applications, the material should be stable and easy to synthesize. These conditions limit the suitable candidate materials, and currently there is urgent need to search for realistic material systems for the realization of the nodal-line and nodal-surface topological semimetal states.

In this work, by using first-principles calculations and symmetry analysis, we show that the Ti$_3$Al family compounds exhibit novel topological band structures with the coexistence of a nodal loop and a nodal surface in the low-energy band structures. Taking Ti$_3$Al as a concrete example, we show that the material has an almost ideal nodal loop in that: (i) the loop is close to the Fermi level; and (ii) it is almost flat in energy with energy variation less than 0.25 meV. Such flatness has not been seen in other proposed materials before. Beside the nodal loop, Ti$_3$Al also hosts a nodal surface in the $k_z=\pi$ plane near the Fermi level, dictated by the nonsymmorphic screw rotational symmetry and the time reversal symmetry. The effect of SOC and discussed and an effective model is constructed to describe the nodal loop. In addition, high-quality Ti$_3$Al have been synthesized and the compound is known to possess other outstanding material properties, such as low density, high specific Young's modulus, high strength, as well as excellent corrosion/oxidation/burn resistances~\cite{add40}. Thus, our result reveals a realistic material platform for the exploration of novel properties of nodal-loop and nodal-surface topological states.

\section{Methods and crystal structure}

To study the electronic properties, we carry out the first-principles calculations based on density functional theory (DFT), using the Vienna ab-initio simulation package~\cite{add41,add42}. The generalized gradient approximation (GGA) with the Perdew-Burke-Ernzerhof (PBE) realization is used for the exchange-correlation potential~\cite{add43}. The projector augmented wave (PAW) method is used to treat the ionic potentials. A plane-wave basis set with a kinetic energy cutoff of 500 eV is employed. 11$\times$11$\times$11 and 15$\times$15$\times$15  $\Gamma$-centered \emph{k}-meshes are adopted for the structural optimization and the self-consistent calculations, respectively. The structures are fully relaxed until the residual forces are less than 0.001 eV/\AA, and the energy convergence criterion is set to be $10^{-7}$ eV. To investigate the surface spectrum, we construct the maximally localized Wannier functions~\cite{add44,add45,add46}, and make use of the WannierTools package~\cite{add47}.

\begin{figure}
\includegraphics[width=8.8cm]{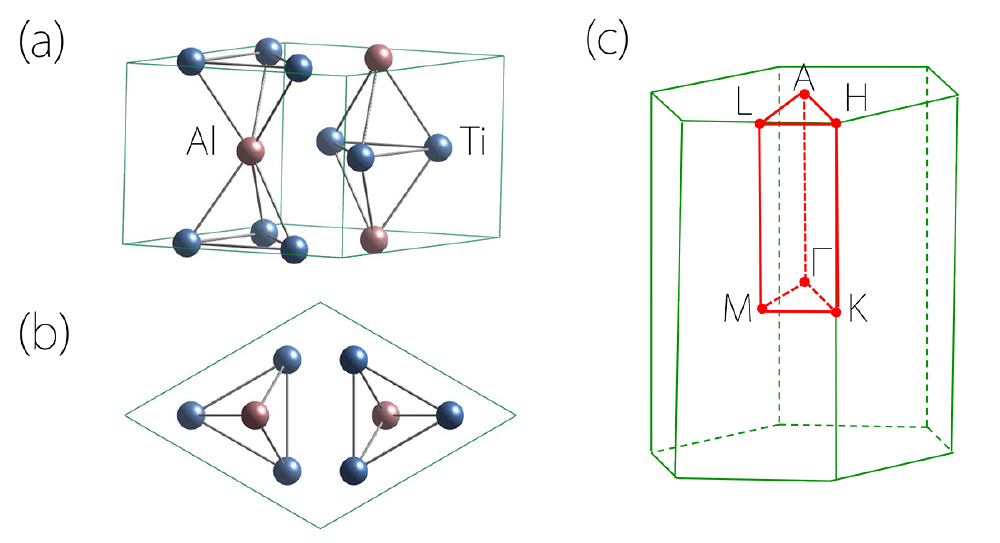}
\caption{Crystal structure of Ti$_3$Al in (a) side view and (b) top view. (c) The Brillouin zone with the high-symmetry points labeled.
\label{fig1}}
\end{figure}

Experimentally, high-quality Ti$_3$Al bulk crystals and nanoparticles have been synthesized by various methods~\cite{add48,add49,add50}. Ti$_3$Al compound naturally crystallizes in the hexagonal \emph{DO$_{19}$} structure~\cite{add48,add49,add50,add51,add52}, with the space group of \emph{P63/mmc} (No. 194). As shown in Fig.~\ref{fig1}(a) and \ref{fig1}(b), in the unit cell of Ti$_3$Al, the bonding between Ti and Al atoms forms two chains of tetrahedrons with a $c/2$ relative shift along the $z$-direction ($c$-axis). The two Al atoms are at the \emph{2c} (1/3, 2/3, 1/4) and the six Ti atoms are at the \emph{6h} (\emph{u}, \emph{1-u}, 1/4) Wyckoff sites, respectively. The optimized lattice parameters are \emph{a} = \emph{b} = 5.756 \AA, \emph{c} = 4.652 \AA, and \emph{u} = 0.8311, which agree well with the experimental values (\emph{a} = \emph{b} = 5.764 \AA, \emph{c} = 4.664 \AA, and \emph{u} = 0.8333)~\cite{add53}. The optimized lattice structure is used in the band structure calculations. Before proceeding, we note that the crystal has the following important symmetries: inversion $\mathcal{P}$, mirror reflection $\mathcal{M}_z$, and two-fold screw rotation $\mathcal{S}_{2z}:(x,y,z)\rightarrow (-x,-y,z+1/2)$. Here, $\mathcal{S}_{2z}$ is a nonsymmorphic symmetry, which involves fractional lattice translations. In addition, no magnetic ordering is observed for the material~\cite{add54}, so the time reversal symmetry $\mathcal{T}$ is also preserved.

\section{Electronic structure}

\begin{figure}
\includegraphics[width=8.8cm]{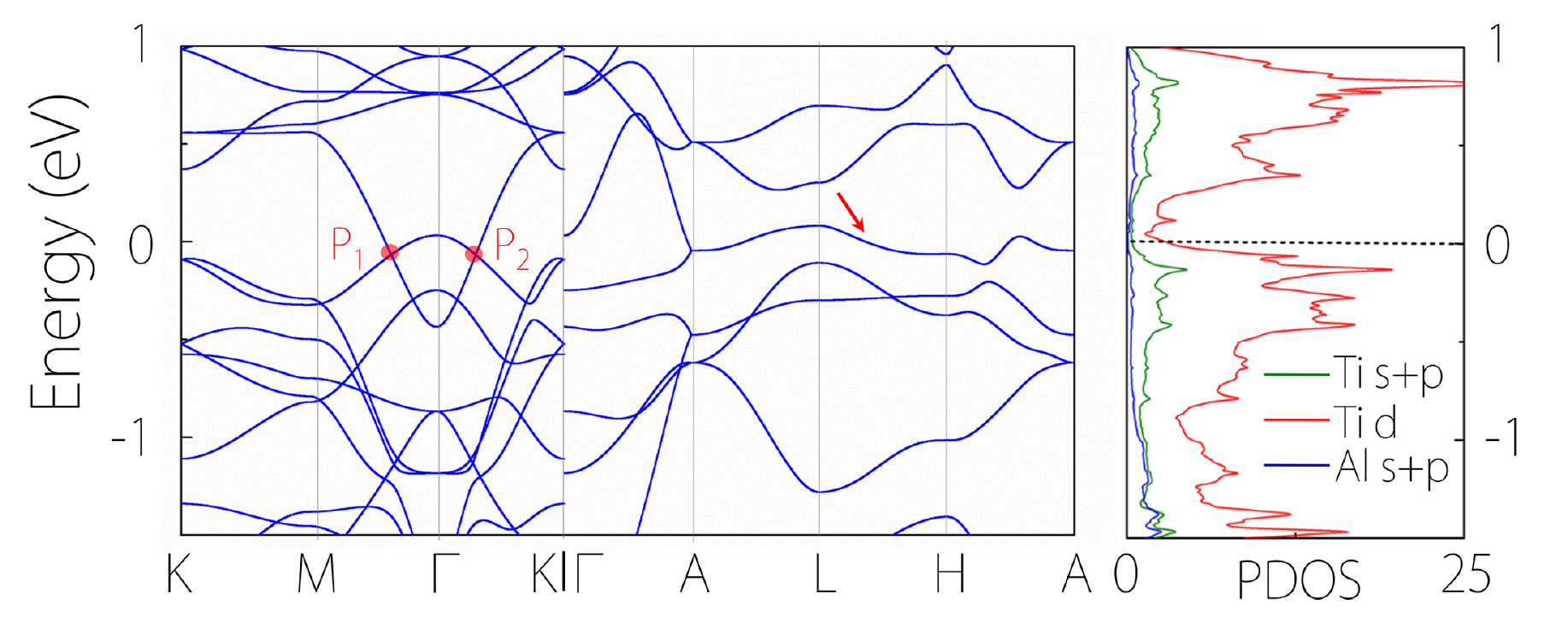}
\caption{Electronic band structure and the projected density of states (PDOS) for Ti$_3$Al. The spin-orbit coupling is neglected. The two band-crossing points labeled as $P_1$ and $P_2$ are two points on a nodal loop centered around $\Gamma$. The arrow indicates the nodal surface in the $k_z=\pi$ plane.
\label{fig2}}
\end{figure}

The electronic band structure and the projected density of states (PDOS) for Ti$_3$Al are plotted in Fig.~\ref{fig2}. Here, SOC is neglected since its strength is very small (due to that only light elements are involved), and its effect will be discussed later. The result in Fig.~\ref{fig2} shows a semimetal feature with suppressed density of states (DOS) at the Fermi level. Besides, there is no spin polarization found in the calculation. These are consistent with the experiment observations~\cite{add54}. From the PDOS plot, we find that the states near the Fermi level are mainly contributed by the \emph{d}-orbitals of the Ti atoms. Notably, there exist several band-crossings in the low-energy band structure near the Fermi level. First, there are two linear band-crossing points near the $\Gamma$ point, which are labeled as $P_1$ and $P_2$ in Fig.~\ref{fig2}. Second, two bands cross linearly at the A point along $\Gamma$-A, and they become degenerate along the paths A-L-H-A. In the following, we shall analyze these two groups of band crossings one by one.

\subsection{Nodal loop}

\begin{figure}
\includegraphics[width=8.8cm]{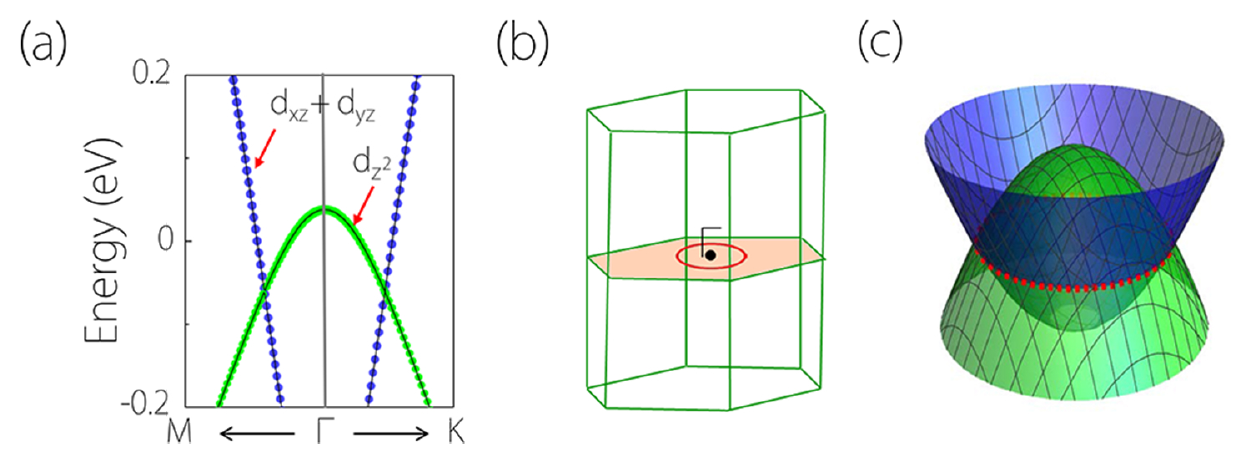}
\caption{(a) Orbital-projected band structure of Ti$_3$Al for paths near the $\Gamma$  point. (b) Schematic illustration of the nodal loop in the Brillouin zone. (c) Dispersion of the bands in the $k_z=0$ plane around $\Gamma$, showing the nodal loop.
\label{fig3}}
\end{figure}

Let's first consider the two linear band-crossing points $P_1$ and $P_2$. Figure~\ref{fig3}(a) shows the enlarged band structure around the two points. One observes that the crossings are associated with the inverted band ordering (between the two crossing bands) around the $\Gamma$ point.
By analyzing the orbital components of the states, we find that one band mainly has the $d_{z^2}$ orbital character, whereas the other band has the $d_{xz}$ and $d_{yz}$ orbital character. Due to the presence of both $\mathcal{P}$ and $\mathcal{T}$ symmetries, such crossing points cannot be isolated~\cite{add12}. We perform a scan of the band structure around the two points and find that the two points in fact sit on a nodal loop in the $k_z=0$ plane centered at the $\Gamma$ point [see Fig.~\ref{fig3}(b)]. In Fig.~\ref{fig3}(c), we plot the band dispersion in the $k_z=0$ plane around $\Gamma$, and one clearly observes the nodal loop formed by the linear crossing between the two bands. This nodal loop is at energy about $-67$ meV, quite close to the Fermi level. Since it is slightly below the Fermi level, it can be readily detected in the angle-resolved photoemission spectroscopy (ARPES) measurement.

\begin{figure}
\includegraphics[width=8.8cm]{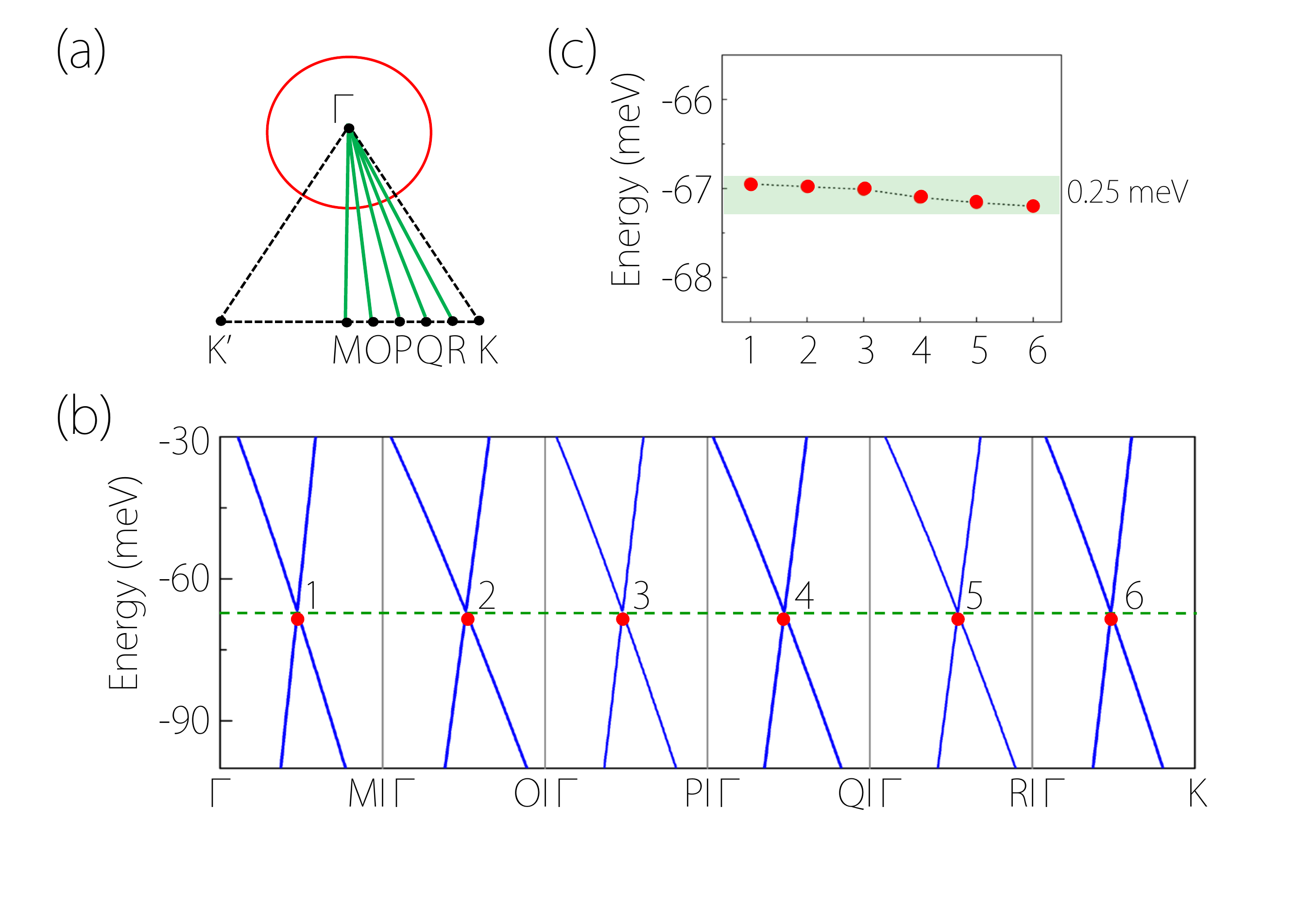}
\caption{(a) Schematic figure showing the paths taken in the $k_z=0$ plane. The red circle denotes the nodal loop. The points O, P, Q, and R are equally spaced between M and K. (b) Band dispersions along the paths indicated in (a). The green horizontal line is at $-65$ meV, serving as a guide of eye. (c) Energy variation for the crossing points in (b). The variation is in a range less than 0.25 meV.
\label{fig4}}
\end{figure}

There is generally finite energy variation along a nodal line (unless there is special symmetry constraint like that for nodal loops in the superconducting excitation spectrum~\cite{add13}). In the proposed nodal-line materials, such variation is typically large, on the order of hundreds of meV. A salient feature of the nodal loop in Ti$_3$Al is that the loop is quite flat in energy. In Fig.~\ref{fig4}(a)-(c), we examine the energy variation for points on the loop. The result shows that the variation is less than 0.25 meV. In comparison, the loop energy variation is $>$200 meV for Cu$_3$PdN~\cite{add16,add17},  $>$300 meV for TiB$_2$~\cite{add24}, and $>$800 meV for the elemental Be crystal~\cite{add20}. Thus, the nodal loop in  Ti$_3$Al can be regarded as exceedingly flat in energy.

As we have mentioned, nodal loops can be classified based on the dispersion type for the points on the loop. A type-I loop is formed by the crossing between an electronlike band and a holelike band; a type-II loop is formed by crossing between two electronlike or two holelike bands; and a hybrid loop typically occurs when one of the crossing bands is saddlelike. Here, from Fig.~\ref{fig3}(c) and Fig.~\ref{fig4}, one easily finds that the nodal loop in Ti$_3$Al is of type-I.

The nodal loop here is robust due to the protection by two independent symmetries in the absence of SOC. The first is the combined $\mathcal{PT}$ symmetry, which requires that the Berry phase along any close path in the system must be quantized in unit of $\pi$. We numerically calculated that for a path encircling the nodal loop, the Berry phase is nontrivial ($\pm \pi$ with the sign depending on the direction of the path), hence the loop cannot be gapped out by weak perturbations that preserve the symmetry. The second protection comes from the $\mathcal{M}_z$ symmetry. The two crossing bands have opposite $\mathcal{M}_z$ eigenvalues in the mirror-invariant plane $k_z=0$ [which can also be easily seen from their orbital characters in Fig.~\ref{fig3}(a)], hence the two bands must cross without hybridization. The $\mathcal{M}_z$ symmetry also confines the loop within the $k_z=0$ plane.

\begin{figure}
\includegraphics[width=8.8cm]{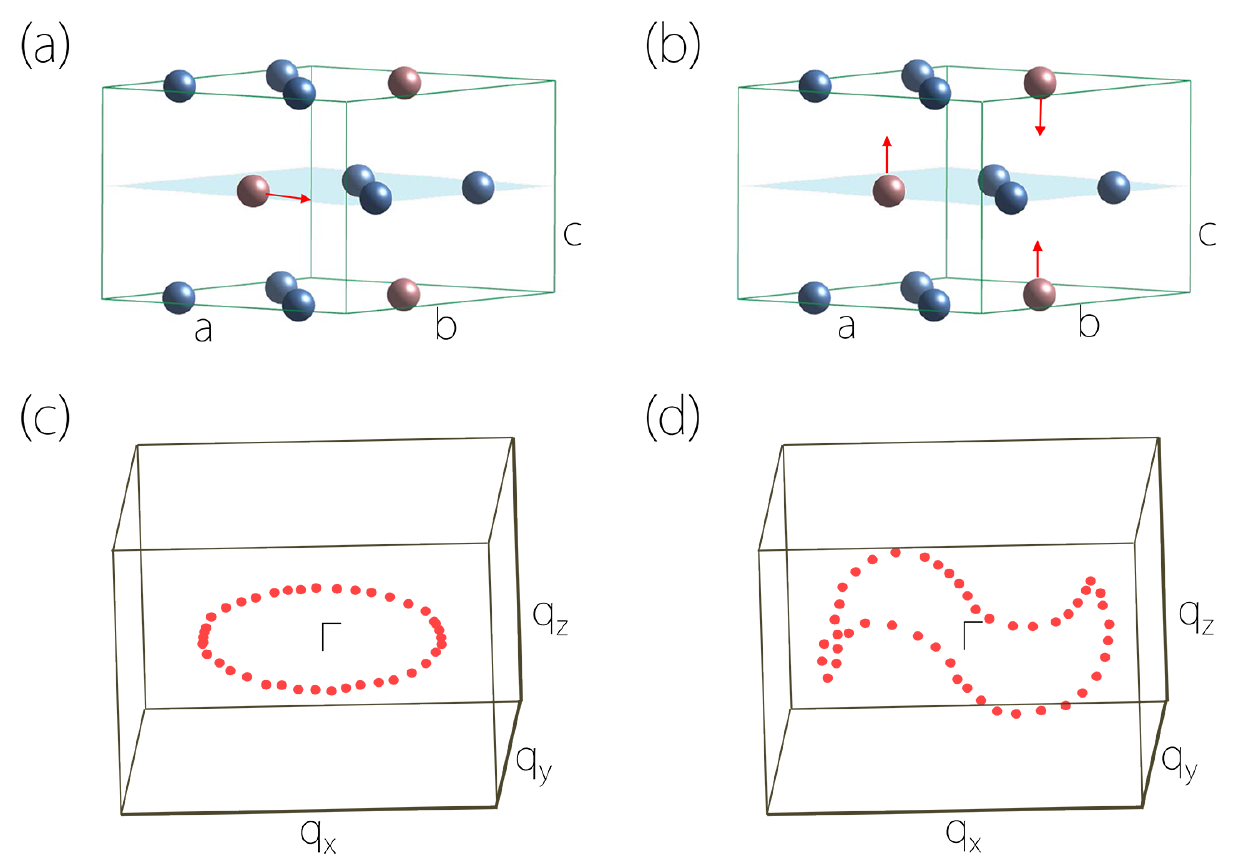}
\caption{(a) Break the inversion symmetry by shifting one of the Al atom in the unit cell along the $a$-axis. Noted, the mirror symmetry of the $k_z=0$ plane is reserved. (b) Break the mirror symmetry of the $k_z=0$ plane by shifting the two Al atoms along the $c$-axis in the opposite directions. In this case, the inversion symmetry is reserved. (c) and (d) are the schematic illustrations of the nodal loop in (a) and (b), respectively.
\label{fig5}}
\end{figure}

It should be noted that the two protections are independent, which means that the loop is robust as long as one of them is preserved. To explicitly demonstrate this point, we consider the following perturbations to the lattice structure. In the first case, we artificially shift one of the Al atom in the unit cell along the $a$-axis, as illustrated in Fig.~\ref{fig5} (a). This breaks $\mathcal{P}$ but preserves $\mathcal{M}_z$. As a result, we find that the nodal loop is preserved and still lies in the $k_z=0$ plane [see Fig.~\ref{fig5} (c)]. In the second case, we artificially shift the two Al atoms along the $c$-axis in the opposite directions [see Fig.~\ref{fig5} (b)], which preserves $\mathcal{P}$ but breaks $\mathcal{M}_z$. One finds that the loop is again preserved, but its shape gets distorted and no longer lies in the $k_z=0$ plane, as shown in Fig.~\ref{fig5} (d). From the discussion, we see that the loop enjoys double protection. Moreover, the $\mathcal{P}$ and $\mathcal{M}_z$ symmetries are not easy to destroy by lattice strains. Such robustness would be advantageous for experimental studies and possible applications.

To further characterize the nodal loop, we construct a low-energy effective model around the $\Gamma$ point. From symmetry analysis, we find that the states at $\Gamma$ of the two crossing bands belong to the \emph{A$_{2u}$} and \emph{A$_{1g}$} irreducible representations of the \emph{D$_{6h}$} point group. Using these two states as basis and with the constraints from the \emph{D$_{6h}$} and $\mathcal{T}$ symmetries)~\cite{add55}, we can construct the $k\cdot p$ Hamiltonian for the two bands around the $\Gamma$ point:
\begin{equation}\label{FNRm}
\mathcal{H}=\left[
              \begin{array}{cc}
                h_{11} & h_{12} \\
                h_{21} & h_{22} \\
              \end{array}
            \right],
\end{equation}
with,
\begin{equation}
h_{ii}=A_{i}(k_{x}^{2}+k_{y}^{2})+B_{i}k_{z}^{2}+M_{i}
\end{equation}
\begin{equation}
h_{12}=-h_{21}=iCk_{z}
\end{equation}

Here, the wave-vector $\bm k$ is measured from $\Gamma$, the expansion is up to $k$-quadratic terms, and the real coefficients $A_i$, $B_i$, $M_i$ ($i=1,2$), and $C$ are material specific parameters. The model describes a nodal loop in the $k_z=0$ plane when $M_1>M_2$, $A_1<0$, and $A_2>0$. From the model, we see that the two bands are decoupled in the $k_z=0$ plane because they have opposite $\mathcal{M}_z$ eigenvalues, and the band ordering is inverted around $\Gamma$, consistent with our previous discussions.
The model parameters can be obtained by fitting the DFT band structure. We find that $A_1 = -9.66$ eV$\cdot${\rm \AA}$^2$, $A_2 = 11.37$ eV$\cdot${\rm \AA}$^2$, $B_1 = 36.22$ eV$\cdot${\rm \AA}$^2$, $B_2 = -25.71$ eV$\cdot${\rm \AA}$^2$, $C = 22.34$ eV$\cdot${\rm \AA}$^2$, $M_1 = 0.12$ eV, $M_2 = -0.52$ eV for Ti$_3$Al, satisfying the conditions above for having a nodal loop.

\subsection{Nodal surface}

\begin{figure}
\includegraphics[width=8.8cm]{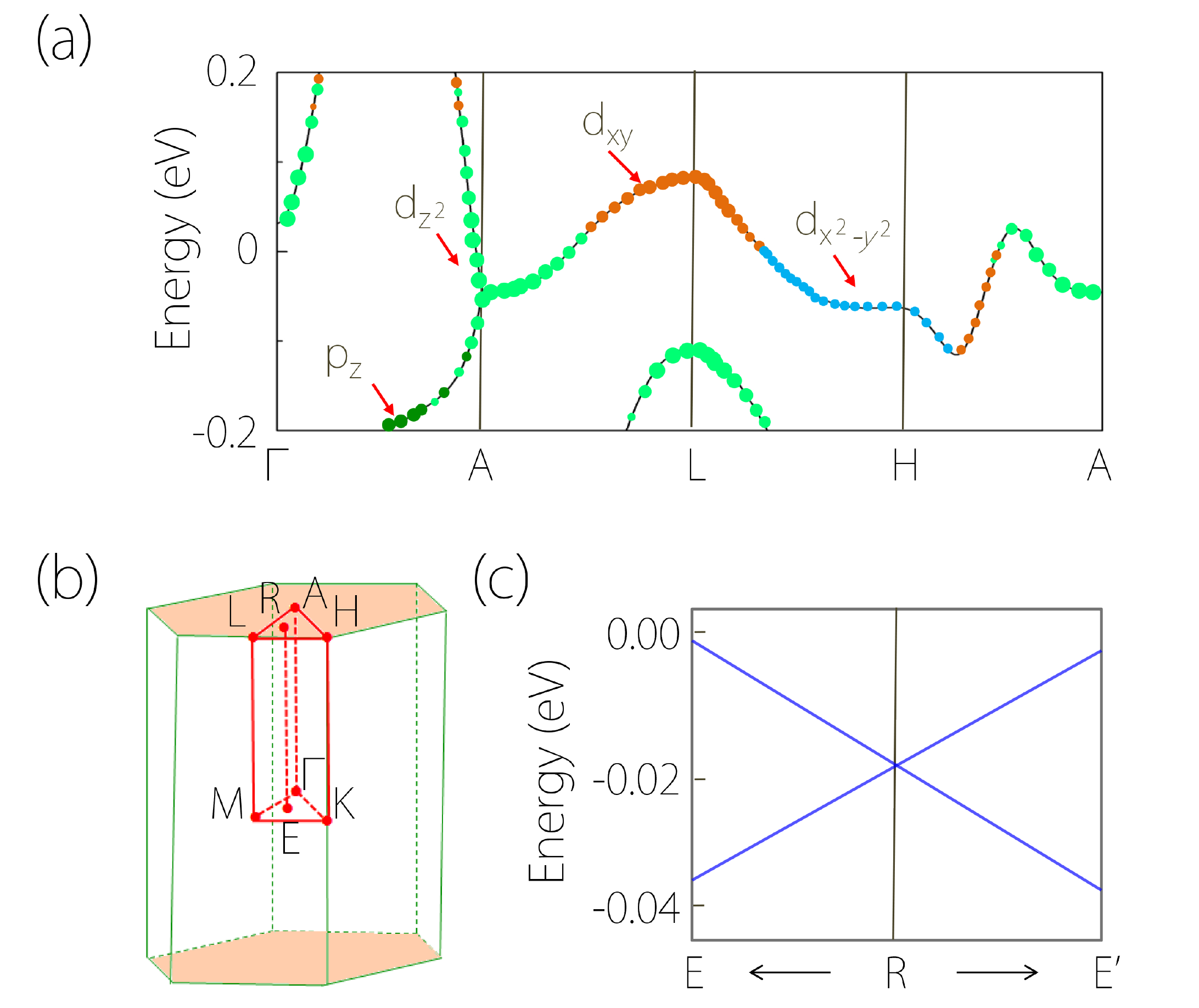}
\caption{(a) Orbital-projected band structure of Ti$_3$Al for $\Gamma$-A and paths in the $k_z=\pi$ plane. ((b) shows the Brillouin zone. R and E are the centers of the triangles A-L-H and $\Gamma$-M-K, respectively. (c) Enlarged view of the low-energy band dispersion along E-R path, showing the linear crossing between two bands along the $k_z$ direction.
\label{fig6}}
\end{figure}

Now let us turn to discuss the degenerate bands along the A-L-H-A path in the $k_z=\pi$ plane. An enlarged view with the orbital projection is shown in Fig.~\ref{fig6}(a). In fact, one checks that all the bands in the $k_z=\pi$ plane are doubly degenerate (without counting spin). And the degeneracy is due to the linear crossing between two bands along the $k_z$ direction (see Fig.~\ref{fig6}(b) and (c) for the crossing at a generic $k$-point in the plane). Thus, there exists a nodal surface in the $k_z=\pi$ plane formed by the crossing between the two bands. As we mentioned earlier, the presence of the nodal surface is dictated by the nonsymmorphic $\mathcal{S}_{2z}$ symmetry and $\mathcal{T}$ symmetry. To understand this point, one notes that in the absence of SOC~\cite{add39},
\begin{equation}
(\mathcal{TS}_{2z})^2=T_{001}=e^{-ik_z},
\end{equation}
where $T_{001}$ is the translation along the $z$-direction by a lattice constant. In the $k_z=\pi$ plane, each $k$-point is invariant under the $\mathcal{TS}_{2z}$ operation, and $(\mathcal{TS}_{2z})^2=-1$ means that there is a Kramers-like degeneracy due to the anti-unitary $\mathcal{TS}_{2z}$ symmetry. Thus, the bands in the $k_z=\pi$ plane must become doubly degenerate, forming the nodal surface. The similar argument has been presented in the previous discussion of the materials BaVS$_3$ and K$_6$YO$_4$~\cite{add38,add39}. Although symmetry requires the presence of the nodal surface and confines its location in the $k_z=\pi$ plane, it puts no constraint on the energy and the dispersion of the surface. Fortunately, the nodal surface in Ti$_3$Al lies close to the Fermi energy and its energy variation is relatively small (see Fig.~\ref{fig6}(a)). These features would facilitate the experimental studies of the nodal surface in Ti$_3$Al.

\section{Discussion and Conclusion}

From the above discussion, we see that Ti$_3$Al hosts both a nodal loop and a nodal surface in its low-energy band structure. Since the loop lies in the $k_z=0$ plane, while the surface is in the $k_z=\pi$ plane, the low-energy quasiparticles around the two, namely the nodal-loop fermions and the nodal-surface fermions would behave almost independently under smooth perturbation, due to their large momentum separation. Hence, for such case, they would contribute separately to physical properties. For example, regarding transport, the nodal-surface fermions should exhibit very strong anisotropy: it is more dispersive along the $z$-direction than in the $xy$-plane (they may be regarded as 1D chiral fermions in the $z$-direction~\cite{add39}). Thus, one can expect that the transport in the $z$-direction is dominated by the nodal-surface fermions (due to the much larger carrier population), whereas the transport in the $xy$-plane would be dominated by the nodal-loop fermions. The magnetic quantum oscillations are also expected to show distinct signatures for the two types of fermions. For a nodal surface relatively flat in energy, the equi-energy surfaces around it would take the shape of sheets normal to $z$ and traversing the BZ. In comparison, the equi-energy surfaces around a type-I loop are toruses. Thus, under a magnetic field, the nodal-loop fermions would form closed cyclotron orbits and contribute to the quantum oscillations (like de Haas-van Alphen oscillation)~\cite{add36}, whereas the nodal-surface fermions would likely form open orbits and not contribute to the oscillation~\cite{add56}. It should also be mentioned that at sharp interfaces or atomic scale defects, scattering with large momentum transfer is possible, which may convert the carriers between the nodal loop and the nodal surface.

We have several remarks before closing. First, in the above discussion, SOC is neglected. The presented symmetry arguments apply in the absence of SOC. The inclusion of SOC will in general gap the nodal ring and the nodal surface. Nevertheless, the opened gap is very small due to the small SOC strength in this material. Our computations indicate that, the SOC gap opened at the nodal loop and the nodal surface is less than 5 meV~\cite{sup}, which is indeed negligible.

Second, nodal loops can also be classified based on how they wind around the BZ~\cite{add27}, which has the topology of a three-torus. Hence, a nodal loop that traverses the BZ is topologically distinct from a loop that does not. The distinction can be characterized by three integers (a $\mathbb{Z}^3$ index), corresponding to the number of times that the loop winds around the BZ in each of the three directions~\cite{add27}. In this sense, the nodal loop here has a $(0,0,0)$ characterization, i.e., it does not traverse the BZ. Consequently, it can in principle be continuously deformed into a point and annihilated without breaking the symmetry of the system. Physically, this corresponds to the process in which the two crossing bands are pulled apart. In comparison, the nodal surface cannot be removed since they are guaranteed by the nonsymmorphic space group symmetry, as we discussed above.

Third, nodal-loop metals often possess nontrivial drumhead-like surface states on the surfaces where the loops have finite surface-projected area~\cite{add12,add13}. For Ti$_3$Al, the nodal loop lies in the $k_z=0$ plane, so the (001) surface should host surface states. However, the nodal surface in the $k_z=\pi$ plane appear in the same energy range as the loop, hence the surface states would largely be buried in the projection of the bulk bands ~\cite{sup}. Thus, the surface states here may not be easily detected in experiment.

Finally, we point out that Ti$_3$Al enjoys other excellent material properties. (i) Ti$_3$Al is easy to synthesize and high-quality samples have been demonstrated~\cite{add48,add49,add50}. (ii) Ti$_3$Al is stable and possesses outstanding mechanical properties~\cite{add40}. The nodal loop and nodal surface features are also found in related materials with the same lattice structure, such as Ti$_3$Ga, Ti$_3$Sn, and Ti$_3$Pb ~\cite{sup}, although for some of the materials, the crossings are away from the Fermi level.

In conclusion, we have proposed that the Ti$_3$Al family compounds are a new type of topological materials. The material has a nodal loop in the $k_z=0$ plane and a nodal surface in the $k_z=\pi$ plane. Both features are close to the Fermi level. Remarkably, the nodal loop is almost flat in energy, with very small energy variation ($<0.25$ meV), much smaller than other nodal-loop materials proposed to date. The nodal loop enjoys a double symmetry protection, by either the $\mathcal{PT}$ symmetry or the $\mathcal{M}_z$ symmetry. The nodal surface is formed by the linear crossing between two low-energy bands in the entire $k_z=\pi$ plane, and is guaranteed by the nonsymmorphic screw rotational symmetry and the $\mathcal{T}$ symmetry. We have constructed an effective model for describing the nodal loop, and discussed the possible physical signatures due to the nodal-loop and nodal-surface fermions. This family of materials also possess other excellent material properties, including the high stability and the excellent mechanical properties, which further make these materials candidates to explore the interesting physics associated with the nodal-loop and nodal-surface metals.

\begin{acknowledgments}
The authors thank S. Wu and D.L. Deng for valuable discussions. This work is supported by the Special Foundation for Theoretical Physics Research Program of China (No. 11747152) and Singapore Ministry of Education Academic Research Fund Tier 2 (MOE2015-T2-2-144). We acknowledge computational support from the Texas Advanced Computing Center and the National Supercomputing Centre Singapore.
\end{acknowledgments}

\end{document}